\documentstyle[jkas]{article}

\beginpage{1}
\endpage{7}
\year{}\volume{}\month{}

\runningauthor {C. PARK ET AL.}
\runningtitle{BETTI NUMBERS OF GAUSSIAN FIELDS}

\month{} \year{} \volume{} \issueno{}
\beginpage{1}\endpage{7}
\date{}

\begin{document}
\title{BETTI NUMBERS OF GAUSSIAN FIELDS}
\author{Changbom Park$^1$, Pratyush Pranav$^2$,
Pravabati Chingangbam$^3$,
Rien van de Weygaert$^2$, \\
Bernard Jones$^{2}$,
Gert Vegter$^4$,
Inkang Kim$^5$,
Johan Hidding$^2$, and
Wojciech A. Hellwing$^{6,7}$,
}
\address{$^1$ School of Physics, Korea Institute for Advanced Study, Seoul 130-722, Korea\\
 {\it E-mail : cbp@kias.re.kr}}
\address{$^2$ Kapteyn Astron. Inst., Univ. of Groningen, PO Box 800, 9700 AV Groningen, The Netherlands}
\address{$^3$ Indian Institute of Astrophysics, Koramangala II Block, Bangalore 560 034, India\\
 {\it E-mail : pravabati@gmail.com}}
\address{$^4$ Johann Bernoulli Inst. for Mathematics and Computer Science,
Univ. of Groningen, P.O. Box 407,\\ 9700 AK Groningen, The Netherlands}
\address{$^5$ School of Mathematics, Korea Institute for Advanced Study, Seoul
130-722, Korea}
\address{$^6$ Institute of Computational Cosmology, Department of Physics,
Durham University, South Road,\\ Durham DH1 3LE, United Kingdom}
\address{$^{7}$ Interdisciplinary Centre for Mathematical and Computational Modeling (ICM), University of Warsaw,\\ ul. Pawinskiego 5a, Warsaw, Poland}

\address{}
\offprints{P. Chingangbam}


\abstract{
We present the relation between the genus in cosmology and the Betti numbers
for excursion sets of three- and two-dimensional smooth Gaussian random fields,
and numerically investigate the Betti numbers as a function of threshold level.
Betti numbers are topological invariants of figures that can be used to distinguish
topological spaces.
In the case of the excursion sets of a three-dimensional field
there are three possibly non-zero Betti numbers; $\beta_0$ is the number of connected regions,
$\beta_1$ is the number of circular holes (i.e., complement of solid tori), and $\beta_2$ is
the number of three-dimensional voids (i.e., complement of three-dimensional excursion regions).
Their sum with alternating signs is the genus of the surface of excursion regions.
It is found that each Betti number has a dominant contribution to the genus
in a specific threshold range.
$\beta_0$ dominates the high-threshold part of the genus curve
measuring the abundance of high density regions (clusters).
$\beta_1$ dominates the genus near the median thresholds which measures
the topology of negatively curved iso-density surfaces, and
$\beta_2$ corresponds to the low-threshold part measuring the void abundance.
We average the Betti number curves
(the Betti numbers as a function of the threshold level)
over many realizations of Gaussian fields and find that
both the amplitude and shape of the Betti number curves
depend on the slope of the power spectrum $n$ in such a way that
their shape becomes broader and
their amplitude drops less steeply than the genus as $n$ decreases.
This behaviour contrasts with the fact that the shape of
the genus curve is fixed for all Gaussian fields regardless of the power spectrum.
Even though the Gaussian Betti number curves should be calculated for each
given power spectrum, we propose to use the Betti numbers for better specification of
the topology of large scale structures in the universe.
}


\maketitle


\section{INTRODUCTION}
The spatial distribution of galaxies is determined
by the primordial density fluctuations and the galaxy formation process.
Gott et al. (1986) proposed to use the topology of the large-scale
structure of the universe to examine whether or not the initial fluctuations were
Gaussian. Since then, a number of studies have focused on measuring the genus of
the large-scale distribution of galaxies and clusters of galaxies, mainly concluding
that the topology of the observed large-scale structure is consistent with
the predictions of the initially Gaussian fluctuation models with the observed
power spectrum and with simple galaxy formation prescriptions
(Gott et al. 1989; Park, Gott \& da Costa 1992; Moore et al. 1992; Vogeley et al. 1994; Rhoads et al. 1994;
Protogeros \& Weinberg 1997; Canavezes et al. 1998; Hikage et al. 2003, 2003;
Canavezes \& Efstathiou 2004; Park et al. 2005; James et al. 2009; Gott et al. 2009).
But the observed topology is found to be inconsistent with the theoretical models
on small scales where the details of galaxy formation prescription become important
(Choi et al. 2010).

The genus is a measure of the connectivity or topology of figures. A full morphological
characterization of spatial patterns requires geometrical descriptors measuring
the content and shape of figures as well as topological ones (Mecke et al. 1994).
Integral geometry supplies such descriptors known as Minkowski functionals.
In a $d$-dimensional space there are $d+1$ Minkowski functionals, and the genus is
basically one of them. Being an intrinsic topology measure, the genus is
relatively insensitive to systematic effects such as the non-linear gravitational
evolution, galaxy biasing, and redshift-space distortion (Park \& Kim 2010).
This is one of the most important
advantages of using the genus in studying the pattern of large-scale structures.

Recently, the study of the topology of large-scale structures of the universe
has been  entering
a new phase with the introduction of the alpha shapes and the Betti numbers
(van de Weygaert et al. 2010, 2011; Sousbie 2011; Sousbie et al. 2011).
In particular, it was realized that the topology of figures can be
specified by the Betti numbers and the genus of the surface of the figures
is only a linear combination of them.
Therefore, Betti numbers are a more versatile and elaborate characterization
of topology.
A goal of introducing these new statistics by recent studies is to avoid
smoothing the observed galaxy distribution in characterizing topology, thereby
increasing the topological information extracted from a given point set.
However, it is often cosmologically useful or required to study the topology
of galaxy distribution on large scales separately where all non-linear effects
are either negligible or can be accurately estimated. This can be done
by smoothing the small-scale structures.

Since the current standard paradigm of cosmic structure formation is
based on an assumption
that the initial conditions are Gaussian, it is essential to know the Betti
numbers of Gaussian fields. While an analytic expression for the genus as a
function of threshold level is known for Gaussian fields, the Gaussian formulae
for the Betti numbers are not known. In this paper we present the numerically
calculated Betti number curves for Gaussian fields with the power-law
power spectra. We then compare the Betti number curves with
the genus curve and discuss their cosmological applications.

\section{BETTI NUMBERS IN THREE DIMENSIONS}
Betti numbers give a quantitative description of topology of figures.
In this section we
will introduce the Betti numbers in three dimensional space, and give
a relation between them and the genus as defined in cosmology.
Suppose we have a Gaussian random field  $\rho({\bf x})$ in a cubic box
with periodic boundaries,
and we measure the topology of the excursion regions (superlevel sets) where
the field value $\rho$ is equal to or above a given threshold ${\bar \rho}+\nu\sigma_0$.
Here ${\bar \rho}$ is the mean and $\sigma_0$ is the rms of the field.
Let us call the set of the excursion regions a space $M$.
Since the Euler characteristic of the three-dimensional excursion regions
and that of their surfaces are related by $2\chi(M) = \chi(\partial M)$ and
the genus satisfies $2-2G_i=\chi(S_i)$
for each component $S_i$ of $\partial M$,
the total genus $G\equiv\sum G_i$ of the boundary $\partial M$ can be written as
\begin{equation}
2N-2G =\sum\chi(S_i) = \chi(\partial M) = 2\chi(M),
\end{equation}
where $N$ is the number of components of $\partial M$.
To be most consistent with previous topology studies in cosmology we use
the following definition for the genus
\begin{equation}
g \equiv G-N = -\chi(\partial M)/2 =  -\chi(M) = -\beta_0+\beta_1-\beta_2,
\end{equation}
where the last equality comes from the Euler-Poincare theorem.
Therefore, the genus is a linear combination of the Betti
numbers with alternating signs.
$\beta_0$ is the number of connected regions,
$\beta_1$ is the number of circular holes, and $\beta_2$ is the number of
three-dimensional voids (see \ref{app:betti} for a mathematical
definition of Betti numbers).

In cosmology the genus of the boundary surface is usually defined as
(Gott et al. 1986)
\begin{equation}
g_c={\rm number\; of\; holes}-{\rm number\; of\; regions},
\end{equation}
where the ``number of regions'' is the number of disconnected
pieces into which the boundary surface is divided and
the ``number of holes'' is the maximum number of cuts that can be
made to the regions without creating a new disconnected region.
The genus defined in this way is not exactly equal to Eq. 1.
When the threshold level is very low and the excursion region encloses isolated
voids, $g = g_c-1$ because $\beta_0=1$.
At very high thresholds they are equal to each other.
When the amplitude of the genus curve is large, this small difference can be neglected.
Gott et al. (1986) actually made an approximation
$g_c = -\chi(\partial M)/2$ and used the Gauss-Bonnet theorem in their
implementation of a computational method of calculating the genus
(Weinberg et al. 1987).

For a given realization of a Gaussian random field we calculate the first
Betti number $\beta_0$ by counting the number of independent connected regions.
$\beta_0$ approaches 0 and 1 in the limit of very large positive and negative
$\nu$, respectively.
On the other hand, the third Betti number $\beta_2$ approaches 0 at both limits.

For Gaussian fields $\beta_0$ and $\beta_2$ as functions of $\nu$ (hereafter Betti
number curves) averaged over many realizations are almost symmetric to each other
when the amplitudes of the curves are large.

\section{NUMERICAL RESULTS}
\begin{figure}[!t]
\center
\epsscale{1.16}
\plotone{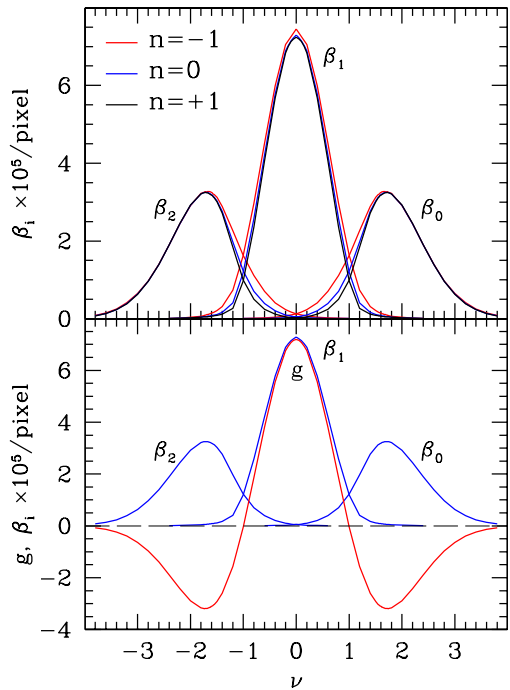}
\vspace{-0.78cm}
\caption{
(upper panel) The Betti numbers per unit pixel volume for three-dimensional
Gaussian fields with
power-law power spectra. Amplitudes of the curves for $n=\pm1$ cases are
scaled to the $n=0$ case according to the scaling law of the genus curve. (lower
panel) The genus and Betti number curves for the $n=0$ power-law power spectrum
case.
}
\end{figure}

We use a $256^3$ array to generate Gaussian random fields
with power-law power spectra smoothed with a Gaussian filter over $R_G=5$ pixels.
$\beta_0$ and $\beta_2$ are calculated by counting the connected super-
and sub-level sets found at each threshold level, respectively.
Then the second Betti number can be calculated from Eq. 2:
\begin{equation}
\beta_1(\nu) = g(\nu)+\beta_0(\nu)+\beta_2(\nu),
\end{equation}
where the genus is calculated by using the code written by Weinberg (1988)
who adopted the method proposed by Gott et al. (1986).

Fig. 1 shows the three Betti numbers of Gaussian fields having power-law
power spectra with power indices of $n=-1, 0,$ and 1.
Each of them is averaged over 100 realizations.
The amplitudes of the Betti number curves for the cases of $n=-1$ and $+1$ are scaled
to the $n=0$ case using the scaling law for the genus curve (see below).
It can be noticed that $\beta_1$ has a maximum at $\nu=0$ and crosses $\beta_0$
and $\beta_2$ near $\nu=+1$ and $-1$, respectively. The $\beta_0$ curve
has a maximum near $\nu=1.7$, which slowly moves to lower
thresholds as $n$ decreases. The overlap between the $\beta_0$ and $\beta_1$
curves or between $\beta_2$ and $\beta_1$ is significant.
There is even an overlap between $\beta_0$ and $\beta_2$.
What is the most important observation is the fact that the shape of the Betti
number curves does depend on the shape of the power spectrum: the Betti number
curves become broader as $n$ decreases. Note that the Gaussian genus curve is
given by (Gott et al. 1986; Hamilton et al. 1986)
\begin{equation}
g(\nu) = -{1\over{8\pi^2}} \left( {{\langle k^2 \rangle}\over 3}\right)^{3/2}
(1 - \nu^2)e^{-\nu^2/2},
\end{equation}
where $\langle k^2\rangle=\sigma^2_1/\sigma^2_0$,
$\sigma_0 = \langle\rho^2\rangle^{1/2}$, and
$\sigma_1 = \langle|\nabla\rho|^2\rangle^{1/2}$.
Its shape is independent of the power spectrum and depends only on
whether or not the field is Gaussian. Only its amplitude depends on the shape
of the power spectrum. The dependence of the Betti number curves on $n$ is
confined in the range of $|\nu|\le 2$.
Being topology measures, the Betti numbers do not depend on
the amplitude of the power spectrum.
\begin{figure}[!!t]
\vspace{-0.145cm}
\centering
\epsfxsize=7.5cm \epsfbox{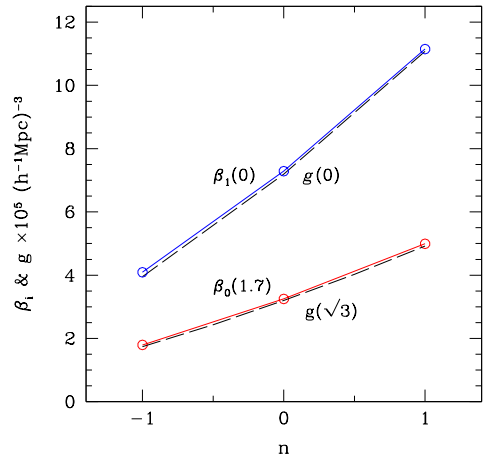}
\vspace{-0.91cm}
\caption{
Amplitudes of the genus and Betti number curves as a function of the spectral
index $n$. $\beta_1$ is compared with the genus at $\nu=0$. The amplitude of
$\beta_0$ at $\nu=1.7$ is compared with that of the genus at $\nu=\sqrt{3}$.
}
\end{figure}

The bottom panel of Fig. 1 compares the Betti number curves with
the genus curve in the case of $n=0$. It can be seen that
the three Betti numbers in three dimensions correspond to
the three extrema of the genus curve.
At large thresholds of $|\nu|>2$ $\beta_0$ and
$\beta_2$ are almost equal to $-g$ because the absolute value of the genus is
very close to the number of excursion sets or voids. The maximum of $\beta_1$ at
$\nu=0$ is a little higher than that of $g$. The difference is larger for
smaller spectral indices as can be noticed in the upper panel.

The amplitude of the Betti number curves depends on the slope of the power
spectrum. Fig. 2 shows the amplitudes of the Betti number curves (solid lines)
together with that of the genus curve (dashed lines) as a function of the
spectral index. The comparison is made at the maxima of the Betti number
and genus curves. As $n$ decreases, the amplitudes of all the Betti number
curves decrease too. The scaling relation is very close to that of the genus
curve, which is $gR_G^3 \propto (n+3)^{3/2}$ (Hamilton, Gott \& Weinberg 1986),
but not exactly the same. $\beta_1(0)$ decreases slower than $g(0)$ as $n$
decreases due to the increasing contributions of $\beta_0$ and $\beta_2$ at $\nu=0$.

The behaviour of $\beta_0(1.7)$ is also very similar to that of $g(\sqrt{3})$
even though the maximum moves slightly to lower threshold values and the
amplitude decreases slower than the genus as $n$ decreases.
The latter behaviour is due to the increasing
contribution of $\beta_1$ near its maximum. Therefore, the amplitudes of the Betti
number curves depend on the slope of the power spectrum near the smoothing
length in a way very similar to that of the genus curve with the difference
also depending on the slope of the power spectrum.

\begin{figure}[!t]
\centering
\epsscale{1.02}
\plotone{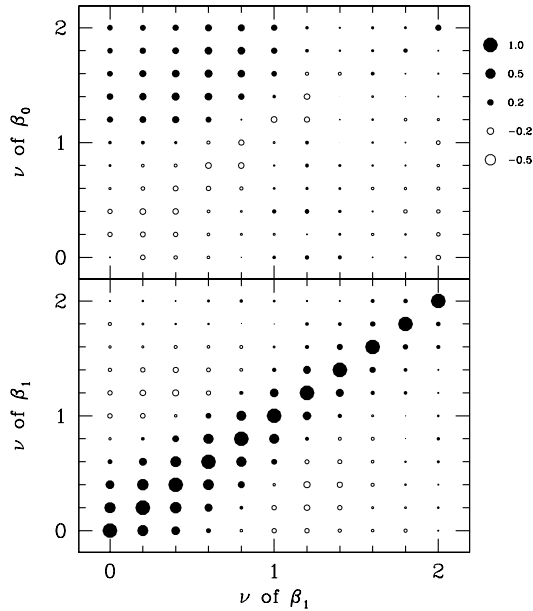}\
\vspace{-0.55cm}
\caption{
(upper panel) Cross-correlations between $\beta_0$ and $\beta_1$ at various
threshold levels. (bottom panel) Corss-correlations of $\beta_1$ between
different threshold levels. The diagonal elements are one by definition.
Results of 1000 realizations are averaged. The magnitude of the correlations
is encoded in the size and color of dots.
}
\end{figure}
\section{CORRELATIONS OF BETTI NUMBERS}
It is interesting to know whether or not the Betti numbers are
correlated with one another and how much correlation the Betti numbers
have between different threshold levels. The upper panel of Fig. 3 show
the cross-correlations between $\beta_0$ and $\beta_1$ defined as
\begin{equation}
C_{ij} = \langle\beta_0(\nu_i)\beta_1(\nu_j)\rangle
/ \sigma_{\beta_{0i}}\sigma_{\beta_{1j}},
\end{equation}
where
$\sigma_{\beta_{0i}}$ is the rms of $\beta_0$ at a threshold $\nu_i$.
It shows that $\beta_0$ near its maximum ($\nu\approx 1.7$) is positively
correlated with $\beta_1$ at low threshold levels. But there are levels where
$\beta_0$ and $\beta_1$ are anti-correlated.
Therefore, the Betti numbers are not independent of one another.

The bottom panel of Fig. 3 shows
the cross-correlations of $\beta_1$ between different threshold levels,
which is similarly defined as above. $\beta_1$ at high threshold levels is almost
uncorrelated with those at other threshold levels, but positive or negative
correlations with nearby levels are seen at low threshold levels.

\section{BETTI NUMBERS IN TWO DIMENSIONS}
There are two possibly non-zero Betti numbers for excursion regions in a
two-dimensional space. Since the main cosmological application of the Betti numbers
in two dimensions will be for the cosmic microwave background (CMB) anisotropy,
we should consider random fields defined on the two-dimensional surface of sphere
$S^2$ or a part of the two-dimensional real space, $R^2$, depending on whether the
sky coverage of the observation is complete or incomplete, respectively.
The first Betti number, $\beta_0$, is the number of
connected excursion regions with $\rho \ge {\bar\rho}+\nu\sigma_0$.
The second one, $\beta_1$, is the number of holes in the connected regions
for fields in $R^2$ or the number of holes minus 1 for fields in $S^2$.

Since the observed CMB sky is always incomplete, we actually consider
the random fields in a part of $R^2$ in the analyses of observed maps.
To be consistent with the previous works in cosmology, we define the genus as
\begin{equation}
g = \beta_0-\beta_1,
\end{equation}
which is equal to $($the number of isolated excursion regions $-$ number of holes in them$)$
(Gott et al. 1990) with the maximum  possible difference of 1.
In $R^2$ this definition is equivalent to Gott et al.'s. In $S^2$, however,
it differs from Gott et al.'s by 1 at low threshold levels because of $\beta_1$.
Since the maximum amplitude of the genus curve measured from observed maps
is typically very large, this difference is not important.
An analytic formula for the genus-threshold level relation
is known in the case of Gaussian random fields (Melott et al. 1989)
\begin{equation}
g = (2\pi)^{3/2}\cdot {{\langle k^2\rangle}\over 2} \nu {\rm exp}(-\nu^2/2).
\end{equation}

We numerically calculate the two-dimensional Betti numbers. A two-dimensional
random field with a power-law power spectrum is generated on a $1000^2$ array
and is smoothed over 10 pixels with a Gaussian filter. For a given threshold
level the excursion regions are found and the number of independent connected
regions are counted to measure $\beta_0$.
Similarly, $\beta_1$ is calculated by counting independent low density regions
surrounded by excursion regions.
$\beta_0$ approaches 0 and 1 at very large and very low thresholds, respectively,
while $\beta_1$ approaches 0 at both limits.
The Betti number curves are averaged over 1000 realizations.
\begin{figure}[!t]
\centering
\epsscale{1.2}
\plotone{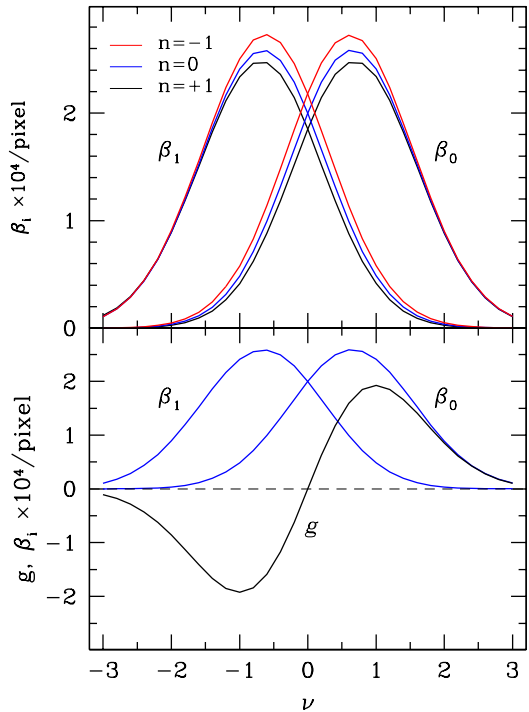}
\vspace{-0.65cm}
\caption{
(upper panel) The Betti numbers per unit pixel area for two-dimensional Gaussian
fields with power-law power spectra. The amplitudes of the curves for $n=\pm1$ cases are
scaled to the $n=0$ case according to the scaling law of the genus curve. (lower
panel) The genus and Betti number curves for the $n=0$ power-law power spectrum field.
}
\end{figure}
The upper panel of Fig. 4 shows the Gaussian Betti number curves for
spectral indices of $n=-1, 0,$ and $+1$. The Betti number curves for $n=-1$
and $+1$ are scaled up and down, respectively, so that their genus curves
at $\nu=1$ match that of the $n=0$ case. It can be seen that the dependence of
the Betti numbers on the slope of the power spectrum is quite large in the range
$|\nu|<1$ and becomes negligible only at high threshold levels ($|\nu|>2$).
The maximum of $\beta_0$ is located around $\nu=0.64$ but this location
shifts slowly towards smaller $\nu$ as $n$ decreases.
When the amplitude of the Betti number curves are very large,
the two Betti number curves are almost symmetric to each other with
respect to $\nu=0$, and their difference has extrema at $\nu=\pm 1$.
We find that $\beta_i R_G^2$ scales approximately like $(n+2+2/3)$.
In comparison the scaling relation of the genus curve is $g R_G^2 \propto (n+2)$.

The bottom panel of Fig. 4 compares the genus curves with the Betti number
curves in the case of $n=0$. Amplitudes of the Betti number curves are higher
than that of the genus curve, and the maxima of $\beta_0$ and $\beta_1$ curves
roughly correspond to the maximum and minimum of the genus curve
located at $\nu=\pm 1$.

\section{DISCUSSION AND CONCLUSIONS}
It should be noted that the three Betti numbers in three dimensions correspond to
the three extrema of the genus curve. $\beta_0$ corresponds mainly to the high-threshold
part of the genus curve that  counts the number of clusters while $\beta_2$
corresponds to the low-threshold part counting the number of voids.
The $\beta_1$ curve is symmetric and corresponds to the median-threshold
part measuring the negative curvature topology of the iso-density contour
surfaces.

Deviations of the genus curve from the Gaussian formula given by Eq. 5
indicate non-Gaussianity.
Likewise, any deviation of the measured Betti numbers from the Gaussian shapes
such as shown in Fig. 1 for the measured power spectrum also signals non-Gaussianity.
Park et al. (1992) and Park et al. (2005) proposed
to use a set of statistics quantifying the deviations of a measured
genus curve from the Gaussian genus curve.
The shift parameter is one of them and is defined by
$\Delta\nu = \int_{-1}^{1}\nu G_{\rm obs}(\nu) d\nu / \int_{-1}^1 G_{\rm fit}(\nu) d\nu$,
where $G_{\rm obs}$ is the measured genus and $G_{\rm fit}$ is the Gaussian curve
best fit to the measured genus data. Therefore, the shift parameter basically
measures how much the central part of the genus curve is shifted from the median
threshold. Since $\beta_1$ dominates the genus near the median threshold,
$\Delta\nu$ basically measures the shift of the $\beta_1$ curve from
the Gaussian expectation.

The high- and low-threshold parts of the genus curves have been
quantified by the cluster- and void- abundance parameters
$A_C = \int_{1.2}^{2.2} G_{\rm obs}(\nu) d\nu / \int_{1.2}^{2.2} G_{\rm fit} d\nu$
and
$A_V = \int_{-2.2}^{-1.2} G_{\rm obs}(\nu) d\nu / \int_{-2.2}^{-1.2} G_{\rm fit} d\nu$, respectively.
They basically measure the amplitudes of the $\beta_0$ and $\beta_2$
curves, respectively, near their maxima relative to those of Gaussian fields
best fit to the measured data. Therefore, even though Betti numbers were not
directly applied to the topology studies by cosmologists in the past, in practice
they have been already used to characterize non-Gaussianity
(see Choi et al. 2010 for an extensive application of these genus-related
statistics to constrain galaxy formation models).

The genus-related statistics have been also used for the genus curve in
two-dimensional spaces (Park et al. 1998; Park et al. 2001).
The shift parameter $\Delta\nu$ is defined similarly to the three-dimensional case.
The asymmetry parameter is the difference ($A_C-A_V$) between the cluster abundance
parameter and the void abundance parameter, where the former is defined
by $A_C=\int_{0.4}^2 g_{\rm obs}d\nu/\int_{0.4}^2 g_{\rm fit} d\nu$, for example.
The parameters $A_C$ and $A_V$ are actually measures of the amplitudes of
the $\beta_0$ and $\beta_1$ curves relatively to the Gaussian cases.

There is quite an inconvenience in using the Betti numbers because of the fact that
both the amplitude and shape of the Gaussian Betti number curves
depend on the shape of the power spectrum and the Gaussian curves
should be calculated for each given power spectrum.
This compares unfavorably with the fact that the shape of the genus curve is
the same for all Gaussian fields regardless of the power spectrum.
However, using all of the Betti number curves to characterize the pattern of
the large-scale structures of the universe is mathematically more justified
than using the genus curve alone since the genus is only a combination of
the Betti numbers and the Betti numbers are also topological quantities.
The genus-related statistics that have been successfully used in previous
works to quantify non-Gaussianities from the genus curve can be now replaced
by the mathematically motivated Betti numbers.
We propose to use the Betti numbers in the future studies of the topology
of large-scale structures.

\acknowledgments{
The authors would like to thank Juhan Kim for help with the numerics,
Bumsik Kim and Marius Cautun for useful discussions and comments, and Herbert Edelsbrunner
for providing us with crucial and incisive insights into topology and homology.
This work has been partially supported by a visitor grant of the
Netherlands Organisation for Scientific Research (NWO).
Inkang Kim acknowledges the support of the KRF grant 0409-20060066.
We thank Korea Institute for Advanced Study for providing computing resources
(KIAS Center for Advanced Computation Linux Cluster System) for this work.}

\appendix
\section{Mathematical Definition of the Betti Numbers} \label{app:betti}
Consider a space $M$ which is a
set of the excursion regions defined by a threshold level $\nu$.
Except for superlevel set with $\nu=-\infty$ or sublevel set with $\nu=\infty$
(in these cases $M=T^3$ and the fourth Betti number $\beta_3=1$),
$\beta_3=0$ since $M$ deformation-retracts
to two-dimensional skeletons or to points. Consequently there are only three
possibly nonzero Betti numbers; $\beta_0={\rm rank}\ H_0(M; Z),
\beta_1={\rm rank}\ H_1(M; Z),$ and $\beta_2={\rm rank}\ H_2(M; Z),$
where $H_i$ is the $i$-th homology group and $Z$ is the integer set
(Munkres 1993).

As the threshold level varies, our space evolves from $\cup B^3$ at very high thresholds
to $T^3\setminus\cup B^3$ at very low thresholds.
In between these limits the space can be $T^3\setminus\cup(D^2\times S^1)$ for example.
If the number of balls or solid tori is very large, the Betti numbers are
approximately those of $R^3\setminus\cup B^3$ and $R^3\setminus\cup(D^2\times S^1)$
in the latter two cases, respectively.
The Betti numbers $(\beta_0, \beta_1, \beta_2)$ in these three cases are then
(number of $B^3, 0, 0), (1,$ number of $D^2\times S^1$,  number of $D^2\times S^1$),
and $(1, 0,$ number of $B^3)$ in decreasing order of $\nu$.
When the space we are considering is $R^3$ and a thick $R^2$ plane cuts
the whole $R^3$ space into two pieces, since $R^2$ is contractible,
the empty plane does not contribute to $\beta_2$.
In the future analyses we will analyze the large-scale galaxy distribution in
a finite volume of the universe with boundaries. In this case we will ignore
the contribution of underdense regions that hit a boundary in calculating $\beta_2$.
However when the space we are considering is $T^3$ as in this work, $R^2$ becomes
$T^2$ and $T^3 - T^2$  becomes $T^2 \times$ interval. Hence
$\beta_2 (T^2 \times$ interval$)=\beta_2(T^2)=1$.
Hence each plane cutting the whole cube contributes 1 to $\beta_2$, and any contiguous
underdense region contributes 1 to $\beta_2$ in $T^3$.







\


{}


\end{document}